\documentclass[rmp,aps,twocolumn,epsfig,eqsecnum,floatfix,amsmath,amssymb]{revtex4}
\usepackage{graphics}
\usepackage{epsfig}
\begin{document}
\def\inbar{\,\vrule height1.5ex width.4pt depth0pt}
\def\IR{\relax{\rm I\kern-.18em R}}
\def\IC{\relax\hbox{$\inbar\kern-.3em{\rm C}$}}
\def\Nc{Noncommutative}
\def\gm{{\bf GM FILL IN}}
\def\nc{noncommutative}
\def\con{conventional} 
\def\Tr{\textrm{Tr}~}
\def\zb{\bar{z}}
\def\eeq{\end{equation}}
\def\beq{\begin{equation}}
\newcommand{\intif}{\int_{-\infty}^{\infty}}
\newcommand{\sid}{\mbox{$\psi^{\dagger}$}}
\newcommand{\sib}{\mbox{$\overline{\psi}$}}
\newcommand{\il}{\int_{-\Lambda}^{\Lambda}}
\newcommand{\ie}{\int_{0}^{\Lambda}}
\newcommand{\iT}{\int_{0}^{2\pi}d\theta}
\newcommand{\iK}{\int_{\Lambda /K_F}^{\pi -\Lambda /K_F}}
\newcommand{\si}[2]{\mbox{$\psi_{#1}(#2)$}}
\newcommand{\etab}{\mbox{\boldmath $\eta $}}
\newcommand{\sigmab}{\mbox{\boldmath $\sigma $}}
\newcommand{\w}{{\omega}}
\def\half{{1\over 2}}
\def\a{{\alpha}}
\def\b{{\beta}}
\def\g{{\gamma}}
\def\d{{\delta}}
\def\t{{\theta}}
\def\s{{\sigma}}
\def\tu{{\tilde u}}
\def\cE{{\cal E}}
\def\cP{{\cal P}}
\def\cT{{\cal T}}
\def\cL{{\cal L}}
\def\cN{{\cal N}}
\def\ve{{\varepsilon}}
\def\e{{\varepsilon}}
\def\ua{\uparrow}
\def\da{\downarrow}
\def\a{{\alpha}}
\def\b{{\beta}}
\def\g{{\gamma}}
\def\d{{\delta}}
\def\L{{\Lambda}}
\def\de{{\varepsilon}}
\def\t{{\theta}}
\def\bK{{\mathbf K}}
\def\bm{{\mathbf m}}
\def\bsig{{\mathbf \sigma}}
\def\bB{{\mathbf B}}
\def\bp{{\mathbf p}}
\def\bI{{\mathbf I}}
\def\bn{{\mathbf n}}
\def\bM{{\mathbf M}}
\def\bq{{\mathbf q}}
\def\br{{\mathbf r}}
\def\bs{{\mathbf s}}
\def\bS{{\mathbf S}}
\def\bQ{{\mathbf Q}}
\def\bs{{\mathbf s}}
\def\bB{{\mathbf B}}
\def\bl{{\mathbf l}}
\def\bPi{{\mathbf \Pi}}
\def\bJ{{\mathbf J}}
\def\bR{{\mathbf R}}
\def\bz{{\mathbf z}}
\def\ba{{\mathbf a}}
\def\bk{{\mathbf k}}
\def\bP{{\mathbf P}}
\def\bg{{\mathbf g}}
\def\bX{{\mathbf X}}
\def\prl{Phys. Rev. Lett.}
\def\prb{Phys. Rev. {\bf B}}
\def\prd{Phys. Rev.{\bf D}}
\def\pre{Phys. Rev.{\bf E}}
\def\rmp{Rev. Mod. Phys.}

\title{Chaotic quantum dots with strongly correlated electrons.}

\author{R.Shankar}
\email{r.shankar@yale.edu} \affiliation{Sloane Physics Laboratory,
Yale University, New Haven, CT 06520}

\begin{abstract}
Quantum dots pose a problem where one must confront three
obstacles: randomness, interactions and finite size. Yet it is
this confluence that allows one to make some theoretical advances
by invoking three theoretical tools: Random Matrix theory (RMT),
the Renormalization Group (RG) and the $1/N$ expansion. Here the
reader is introduced to these  techniques and shown how they may
be combined  to answer a set of questions pertaining to quantum
dots.
\end{abstract}

\maketitle \tableofcontents

\section{INTRODUCTION}
\label{sec:intro}

This colloquium is based on a lecture entitled "Dots for Dummies" I have
frequently given.  The title was chosen, not to offend, but to keep  experts in the audience  from hijacking the lecture with minutiae while
the intended goal was to introduce certain problems involving  quantum dots to a broad audience
not necessarily working in this subfield. The present article attempts to do the same for the general readership of this journal. To follow this colloquium you must be familiar with the rudiments
of second quantization and Feynman diagrams.

The emphasis of this article is idiosyncratic. I am partial to
what I know best and like to talk about most: a frankly
pedagogical tour of several useful techniques and their
applications to this problem. There is not a comparable  emphasis
on phenomenology, for which  I will direct you to other excellent sources.

So let us begin by asking   "What are some of  the questions  in
the world of quantum dots, what are some of the tools brought to
bear in addressing them, and what are some of the answers?"

For our purposes, the dot is an island of size $L$ (in the
 nanoscale  ) within which electrons are restricted to  live.

 If the dot is very small and the level spacing large (compared to other scales like temperature and  interaction strength) we may be interested in just a few quantum states of the dot. For example if the dot is essentially a two-state system, one may study it with the intention of using it as a qubit in quantum computation. Here one needs to understand in detail the
 level structure of {\em a particular dot } so we can manipulate (program) it in a controlled way.

The focus here is on  large dots with a very fine level spacing. These dots are closer to the bulk system of interacting fermions in that a large number of energy levels are in play. However the finite size and finite level spacing are essential features. These dots are very much like nuclei- finite  systems of interacting fermions-  with the main difference that the confining potential is externally imposed and not internally generated. The hard-wall boundary of the dot
 is assumed to be sufficiently irregular so that classical motion is chaotic at and around the
 Fermi energy. The dot is otherwise dirt-free and motion within is ballistic. The system is assumed to be quantum-coherent across the length of the dot.

\begin{figure}[b]
\epsfxsize=3in  \epsfbox{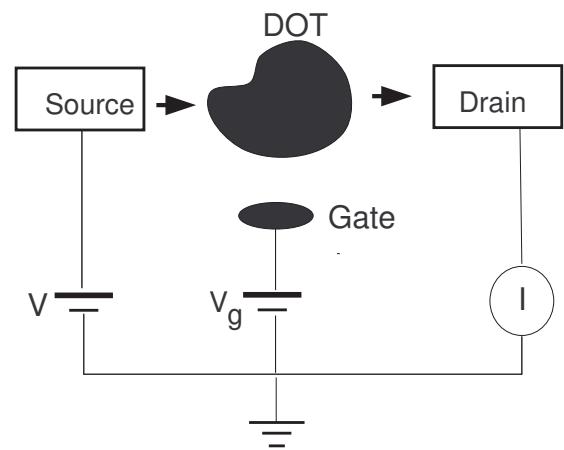} \caption{A typical situation
where leads bring in electrons which tunnel into and out of dot (opposite to the current shown by arrows). A
gate voltage $V_g$ is the control parameter and $V$ is the vanishingly small source-drain voltage difference.\label{setup}}
\end{figure}

 The dot is juxtaposed with leads that act as source and drain for electrons. Electrons are allowed to tunnel in and out of the
 dot. A gate voltage $V_g$ allows one to vary the dot energy relative to the Fermi energy of electrons
 in the  leads as in Figure (\ref{setup}). A tiny voltage $V$ is applied between the source and drain  to drive the
  current, and
  the conductance $G= I/V$  (in the limit of vanishing $I$ and $V$) is measured as a function of the gate
 voltage $V_g$ sketched in Fig (\ref{G}).\footnote{ There is a proportionality factor relating the actual gate voltage to the  applied gate voltage which we have set equal to  unity.}

\begin{figure}
\epsfxsize=3in  \epsfbox{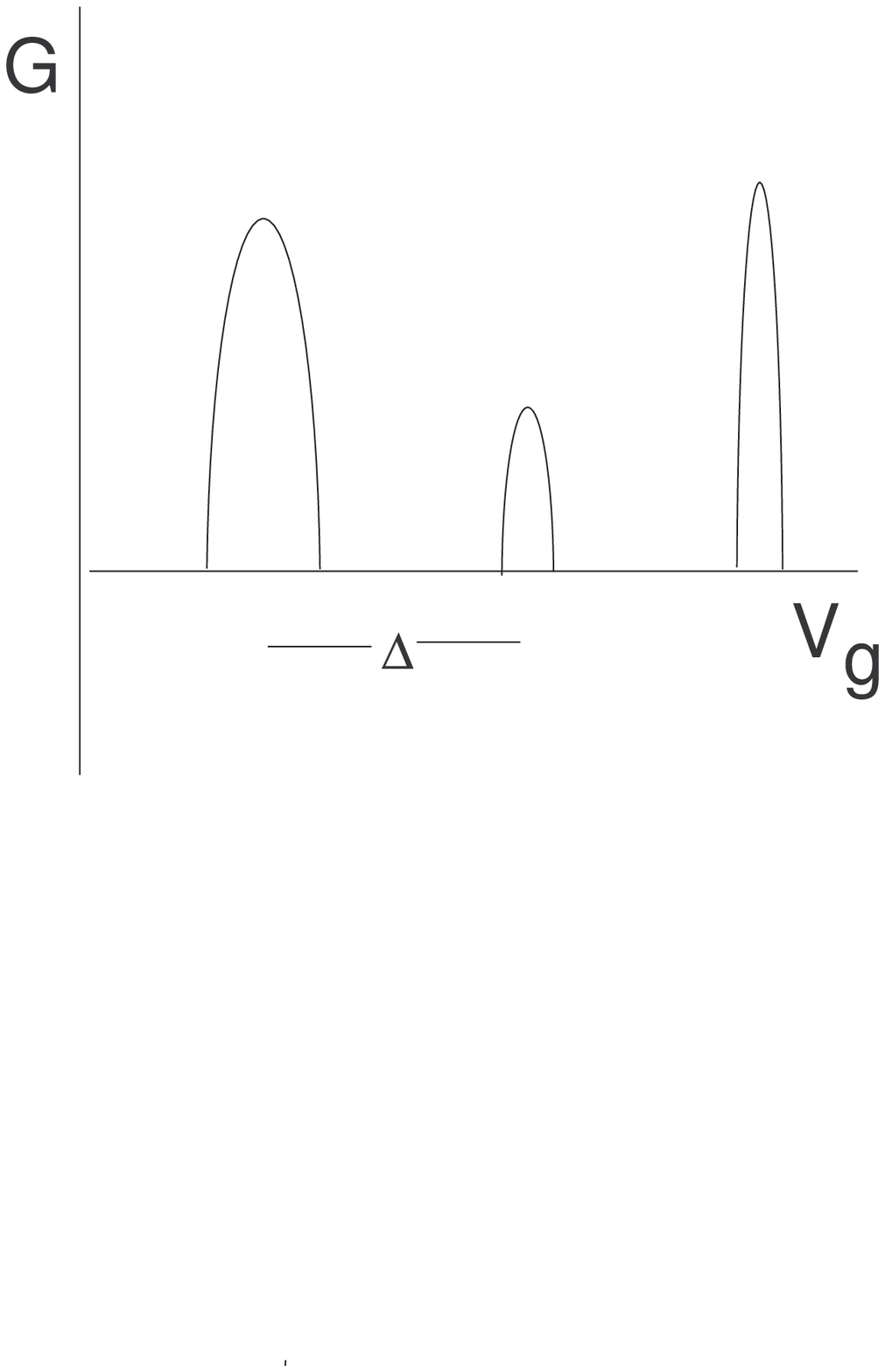} \caption{A caricature of
conductance $G$ versus gate voltage $V_g$. The peaks have varying
heights, widths and spacings.\label{G}}
\end{figure}

 The theoretical challenge is to describe the observed series of peak positions
 and peak heights \cite{qd-reviews1,qd-reviews2,qd-reviews3,H_U-reviews1,peak-height-th,peak-height-expt,peak-height-expt2}.
 These are decided by the shape of the empty dot, coupling to the leads, the number of electrons in it and the interactions between them.    Now, no one disputes that given enough such information,
   every peak may be described in greatest detail
 since the underlying laws (nonrelativistic quantum mechanics and the Coulomb interaction) are thoroughly understood. However this is not the goal here. The goal is more akin to that set by Wigner  in describing excited states of nuclei.  A nucleus, like a dot, is a finite many-body system  with strong interactions between its constituents. Wigner argued that while one could (in principle)  describe the specific levels of a particular nucleus with its  incredibly complicated Hamiltonian, it would more interesting to describe {\em statistical } properties of the spectrum. More specifically he suggested the following program. Consider a {\em mathematical }  ensemble of Hamiltonians which have the same symmetries as the nuclear Hamiltonian (say, being  real and hermitian) and have the same average level spacing. Calculate the statistics of level spacings in this ensemble. Compare this to the actual levels by averaging over members of the {\em physical } ensemble. \footnote{It can sometimes  happen that    any large segment of the spectrum of almost every member of the physical ensemble exhibits these statistical trends, in which case we can look at just one member.}

 Note the difference with Statistical Mechanics, as emphasized  by Guhr {\em et al}  \cite{qd-reviews1}. There  we consider an {\em ensemble of systems} with the same Hamiltonian but different initial conditions, while here we  consider  an {\em ensemble of Hamiltonians} with the same symmetry properties as the given nuclear Hamiltonian.

   In the case of  dots,
   the members of the physical ensemble contain dots with the same density of states. One can generate members of the  ensemble by
 varying the number of electrons in one dot (which probes different energy regimes)  or by manufacturing many similar dots of the same area. If dots of different sizes are used, one can rescale their energies so that the mean spacing is the same. The mathematical ensemble consists of Hamiltonians of the same symmetry class (e.g., real) and same average level spacing. In cases where there is a magnetic field (and the ensemble contains complex unitary Hamiltonians) one can also vary the  field to move around the ensemble. (A minimum change in field may be needed to generate an independent member.)

 Here are two kinds of questions one could ask:
 \begin{itemize}
 \item If I measure  $\Delta$, the difference in gate voltage between
 successive peaks,  what will be the probability distribution $P(\Delta
  )$?
 \item If I measure  $G$, the maximum of each conductance  peak,  what will be
  the probability distribution $P(G
 )$?
 \end{itemize}

 There are of course other issues that we could discuss, but will not, such as the
   width of each peak as a function of
 temperature or spin content.

To get a feeling for the results, let us see how we would go about
explaining the data,  armed with just the theory of
non-interacting electrons (whose spin will be initially ignored.)

To this end let us consider a {\em particular} dot of some
particular shape. We would naturally begin by solving the
single-particle Schr\"{o}dinger  equation
 \beq
 H_0 \phi_{\a} (\br ) = \ve_{\a} \phi_{\a} (\br )\label{SE}
 \eeq
 with  $\phi_{\a} (\br )=0$ at the boundary. This would have to be
 done on a computer for a generic dot. Imagine that we have all
 the wave functions $\phi_{\a}$ and energies $\ve_{\a}$. Let
 $\delta_{i}=\ve_{i+1}-\ve_i$ be the  spacing between levels
 $i$ and $i+1$ and let $\delta$ stand for a generic one, as well as the average level spacing
 (which is the inverse density of states).

 In the non-interacting theory, it is easy to see that at a randomly chosen value of gate voltage $V_g$, there
 will be no conduction. This is because the Fermi energy of the
 electrons in the lead will typically lie between occupied and empty levels
 in the dot, so that transmission by  the occupied levels is forbidden by the Pauli
 principle and transmission by the empty levels is forbidden by energy conservation. However, when the Fermi
 energy of the leads equals an energy level of the  dot there will
 be  transmission. Note that  during transmission the dot has
 the same  energy with N or  N+1 electrons. This feature survives
 even when interactions are
 included.

 In this free-particle model, $\Delta_{i,i+1} $, the difference in the gate voltage $V_g$
 between  peak $i+1$ and
 peak $i$ times $-e$, the charge of the electron,
 will equal $\delta_{i,i+1} =\ve_{i+1}-\ve_i$,  the  difference in energy
 between the level that was responsible for
  peak $i+1$ and the one that was responsible  for peak $i$. It
  follows that (suppressing the constant $-e$)
  \beq
 P(\Delta ) = P(\delta ).\eeq
 In other words, the distribution of spacings in $V_g$ between successive peaks  is the same as
 the distribution of  level spacings in the dot. Consequently, to
  obtain the statistical distribution $P (\Delta )$ theoretically, we need to solve for a
 large number of energy levels in one particular  dot, record the
 spacings $\delta $, and repeat within the ensemble, which means here   other dots with similar density of states
 but arbitrary shape. (In order to fold in the results from many dots and many
 energy ranges, it will be   necessary to rescale energies  so that
 $\delta$, the average level spacing, is fixed at some value.)

 As for the height of any conductance peak, it will be given by
 the product of the probabilities for hopping on to the dot at the
 left lead $L$,  and hopping off on the right lead $R$. These are determined
  $\phi_{\a} (L/R)$, the value near
  the leads of the  wave function $\phi_{\a}$
  corresponding to the state which is responsible for the transmission.  One could collect the statistics of these as well,
  from our numerical solutions to Eqn. (\ref{SE}).

Now it turns out we can spare ourselves a lot of trouble in
obtaining these probability distributions $P(\delta )$ and $P(G)$
by appealing to Random Matrix Theory or RMT, provided certain
condition apply. So we begin with a crash course on RMT.

The first step is to acquaint ourselves with $E_T= \hbar v_F/L$,
the Thouless energy, where $v_F$ is the Fermi velocity. Evidently
$E_T$ stands for the uncertainty in energy of an electron that
traverses the dot ballistically at the Fermi velocity in a time $\tau = L/v_F$. Thus if the
dot
 is connected to big fat leads,
 \beq
 g\simeq {E_T \over \delta}\eeq
 single particle levels in this energy window will each contribute a maximum of $e^2/h$ to conductance $G$. Thus in this case of a dot connected strongly to leads,
  $g$ will be the dimensional conductance, i.e., $G$ measured in units of the quantum of conductance $e^2/h$. .

  While this is one way to introduce $E_T$, you could object that in the experiments we
 consider the leads are weakly coupled to the dot and the levels are sharp.
 You would be right, and it turns out we are interested in $E_T$ for the following different reason.

 If the dot has a  generic shape with no  conserved quantities  except for
 energy,  and the classical dynamics at the Fermi energy is  chaotic,
   {\em the statistics of energy levels lying within
    a band of width $E_T$ and of the corresponding wave functions,
 are determined by RMT\cite{RMT}, given just general features like the average level spacing
 $\delta$ and symmetries of the Hamiltonian.} This is like saying
 that the  statistical properties of an isolated  box of gas are determined by just
  gross features like
 volume, number of particles  and energy. Such a statistical  approach to the energy levels of very complicated systems was first taken by Wigner in the case of nuclei, which are once again finite systems made up of a large number of particles subject to strong interactions that defy any direct treatment. Notice that to use RMT you just need to know the symmetry class (real or complex matrix elements etc.) but no further details including even the spatial dimensionality of the problem.

The first  RMT prediction is   that $P(\delta )$,  the
distribution of exact energy
 level differences
 $\varepsilon_{i+1}-\ve_i$ (for states lying with a band of width
 $E_T$),
 will be indistinguishable from that of an ensemble of  {\em random matrices} of the same
 symmetry class, which in our problem,  consists of all real
 hermitian matrices with the same  $\delta$,  the average level spacing.
 (For a dot in a magnetic field, where time-reversal symmetry is
 broken, the ensemble will consist of complex hermitian matrices.)

 In other words,  the  results
 one person gets by  painstakingly solving  for the spectrum
of an ensemble of dots and compiling the level statistics is same as that compiled by another person
 who picks real hermitian Hamiltonians out of a hat, diagonalizes them   and
plots their spacing distribution! Furthermore this universal
distribution is known in analytic form as soon as we provide
$\delta$.

So, history repeats itself: when  things get really bad, (the
dynamics goes from integrable to chaotic) they become good again
because statistical methods become applicable.

 The second RMT result  bears on wave functions. A common
  quantity of  interest  is the ensemble average (denoted by
  $\langle \cdots \rangle$) of products of wave functions such as
 \beq
 \langle \phi_{\a}^{*}(\br )\phi_{\beta}(\br' ) \rangle
 \label{ficor}
 \eeq
 defined as follows.
  Pick one realization of the dot
 and
 number the states by energy, with $\a$ and $\beta$ being two such numerical
 labels. Pick two points $\br$ and $\br'$ inside the dot. Find $\phi_{\a}^{*}(\br )\phi_{\beta}(\br' ) $. Now
 change the dot smoothly, staying within the ensemble. Since there
 is no level crossing in a chaotic dot, the labels $\a$ and
 $\beta$ retain their integrity. Now recompute $\phi_{\a}^{*}(\br )\phi_{\beta}(\br' )
 $. Find the average of such terms over the entire ensemble. This
 is what Eq. (\ref{ficor}) means. Such correlations are  of use in computing
 peak-height distributions \cite{peak-height-th}.

 In our problem  we need the
 correlation of
  wave functions in momentum space rather
 than coordinate space and we shall use them     to  deal with electron-electron
 interactions rather than  to calculate $P(G)$.

\begin{figure}[t]
 \epsfxsize=3.2in\epsfysize=4.4in \hskip
0.0in\epsfbox{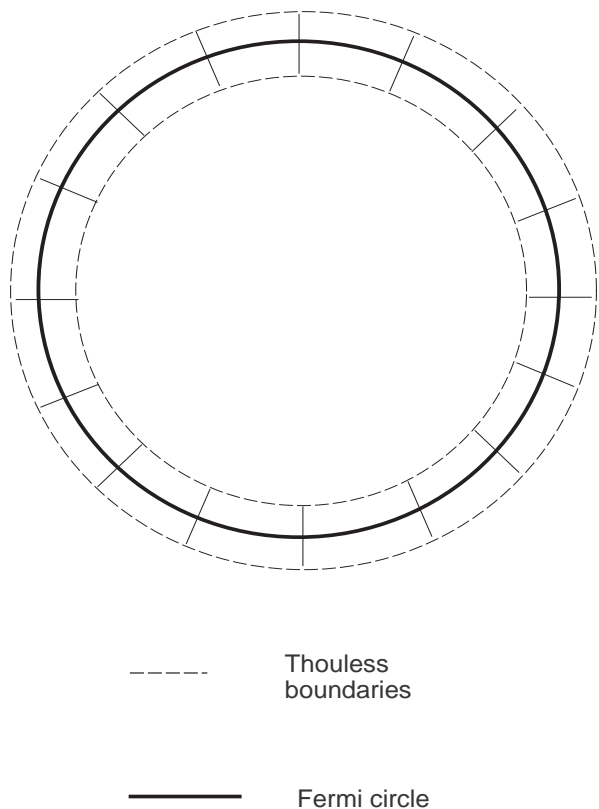} \vskip 0.15in \caption{The Wheel-of-fortune
states within a band of energy $E_T$ concentric with the Fermi
circle. There are roughly $g$ such states of mean momenta $\bk$
centered on $g$ equally spaced points on the Fermi circle. The WOF
states are obtained by chopping off plane waves of the desired
mean momentum at the edges of the dot. These states are very
nearly orthonormal.}
\label{wof}       
\end{figure}

  Let us begin by modifying  the  definition of
 "momentum space" as appropriate to the dot.

 Consider a circular Fermi sea in $\bk$ space
 and a concentric  annulus of width $E_T$ in energy. In the bulk, this region
 contains an infinite number of $\bk$  states. If we now go to the dot of
 size $L$, the best we can do {\em vis-a-vis} momentum is wave packets centered at some $\bk$
 and of width $1/L$ in both directions. It is readily verified that
 we can form  $g$ such "Wheel-of-fortune" (WOF) states (as in Figure \ref{wof})) within
 this annulus \cite{billiard}. Suppose we expand  the $g$ exact eigenstates (labeled by $\a$) of a dot
 within the Thouless band
 in
 this WOF basis (labeled by $\bk$) via the functions $\phi_{\alpha}(\bk )$ as follows
 \beq
 |\a \rangle = \sum_{\bk} \phi_{\alpha}(\bk ) |\bk\rangle .
 \eeq
 Then a typical  RMT result invoked  here is that as $g \to \infty$
 \beq
 \left< \phi_{\alpha}^{*}(\bk ) \phi_{\beta}(\bk' )\right> ={\delta_{\bk
 \bk'}\delta_{\alpha \beta}\over g} + {\cal O} ({1 \over g^2}) \label{correlation}
 \eeq
where the $<\cdots >$ denote describe an average over an ensemble
of similar dots.

Note that Eqn. (\ref{correlation})  is the minimal correlation we
must have: in each sample,  $\alpha$ and $\bk$ label two
orthonormal bases, so that if we set $\bk =\bk'$  and sum over
$\bk$ we must get $\delta_{\alpha \beta}$, {\em sample by sample}
and hence on average as well. (The same goes for setting $\alpha
=\beta$ and summing over them to get $\delta_{\bk \bk'}$.) Similar
correlators exist for products of four wave functions and these
have the form of Wick's theorem.

Unlike the result on level spacings, which can be
demonstrated analytically, this one on wave functions is an
assumption we will make. A reader armed with the requisite
determination and computing power is invited to confirm or
contradict this assumption. All we can say is that several
consequences of this assumption have been verified in our
numerical work  \cite{gang4}.

Before proceeding let me address a common question. Is there any
reason to believe that the $g$ exact eigenstates within $E_T$ can
be expanded in terms of the $g$ WOF states? We \cite{billiard} have
verified  the following in a numerical study of a
"billiard", or dot. First we manufactured the $g$ states of mean
momentum $\bk$ by choosing $g$ equally spaced points on the Fermi
circle and then chopping off the plane waves of these momenta   at
the edges of the billiard.  (We chose  $g=37$ in our study.) Then
we verified that these wavefunctions were orthonormal to an
excellent accuracy. Next we asked how good a basis these functions
formed for expanding the $\alpha$ states within $E_T$. We found
the exact eigenstate at the middle of the Thouless band, i.e, at
the Fermi energy, retained  more than 99.9\% of its norm upon
projection to the WOF basis. As we left the center of the band the
overlap decreased and dropped to around 50\% at the edges. Thus
our results get more  reliable as we go deeper into
 the Thouless band centered at the Fermi energy.

Let us turn to  the data on $P(\Delta )$, the distribution of
spacings in $V_g$ between peaks. In the noninteracting case we saw
that $P(\Delta )= P(\delta )$, and then  we saw that $P(\delta )$
was given by RMT.

 So let us  test the RMT results for level spacings  $P(\delta
)$, against   $P(\Delta )$, the measured distribution of spacings
in $V_g$ between successive peaks. These will match only if the
assumptions of non-interacting electrons is correct.

The test fails. It is found that the peak spacings are an order of
magnitude larger than level spacings in the dot. The reason is of
course electron-electron interactions. Even if we ignore the full
quantum mechanical treatment of interactions, we need to
acknowledge the classical fact that when we add an electron to a
dot, it experiences Coulomb repulsion for the other occupants.
Thus the energy cost of adding an electron is not just that of
going to the next empty level, it is the additional repulsive
energy. This capacitive charging energy is proportional to $N^2$
and we must add a term $u_0N^2$ to the second quantized
Hamiltonian. By studying the data one can come up with a good fit
to $u_0$.

However we can anticipate that this alone will not work in the
presence of spin, even in the non-interacting case. With spin
present, each single particle orbital can be doubly occupied. Thus
adding an electron will (will not) cost us a single particle
energy level difference if it brings the total occupancy to  an
odd (even) integer. Combined with the above mentioned charging
energy, we expect the peak spacing to have a bi-modal distribution
as N varies from odd to even.

This is however not seen. The reason is  based on
the exchange interaction. While it makes sense to put two
electrons in each orbital if one is minimizing  kinetic energy,
this may be a bad idea in terms of potential energy:   the
electrons in any given orbital would have opposite spins and hence
the Pauli principle would not keep them away from each other, thereby raising the Coulomb
repulsion. To minimize the potential energy we must occupy each
orbital once and place the electrons in the same spin state so
that the Pauli principle can keep them apart in space. To let the system
decide what is best we must give it an incentive for higher spins
 by adding a term $-JS^2$, where $S$ is the total spin.

Thus we are led to the following (second-quantized) "Universal" Hamiltonian:
\cite{H_U-reviews1,H_U-reviews2,universal1,universal2,universal3,H_U-kurland}\beq
H_U=\sum_{\alpha}\psi^{\dag}_{\alpha}
\psi_{\alpha}\varepsilon_{\a}+u_0 N^2 -J_0S^2, \eeq where  $\psi_{\a}$ destroys
an electron in a state $\a$. A  third possible
interaction term pertaining to superconducting fluctuations has
been dropped on the grounds that it is unimportant for the dots in
question.

Some proponents of the universal Hamiltonian give the following
argument for why  no other interactions need be considered.
Suppose we take any other familiar interaction and transcribe it
to the exact basis. Sums of products of  random wavefunctions
$\phi_{\alpha}$ will appear and lead to terms with wildly
fluctuating  phases. These terms can be shown to have zero
ensemble average as $g \to \infty$.  Since deviations from the
zero average will be down by $1/g$ we can drop them at large $g$.
By contrast the two terms kept (which commute with $H_0$) are free
of oscillations and survive ensemble averaging.

While  the  success of $H_U$ in explaining a lot of data is
unquestioned,  the accompanying arguments are not persuasive. In
particular,  ensemble averages should be performed not on the
Hamiltonian but on calculated observables. It is also not clear that
a  group of terms should be dropped because they are small, since
they could ultimately prove important to physics at very low energies (low,
even within the already tiny band of width $E_T$).

Now there is a tried and tested way for determining the relative
importance of possible interactions in the low energy limit.  It
is  the
 Renormalization Group (RG). Not only can it    tell us how important any given  interactions is in the low energy sector,
  if there are competing interactions, it can provide an
{\em unbiased} answer in which they  are all allowed to fight it out. (As explained in \cite{rmp}, The
Luttinger Liquid in $d=1$ is an example in which Superconductivity
and Charge Density Wave formation compete and actually annul each
other while in $d=2$ the latter wins at half-filling on a square
lattice.)

 It is therefore natural to ask if the  RG applicable in this problem and if so what
 its verdict is.

\section{The Renormalization Group - the ten cent tour}

In order not to leave behind readers unfamiliar with the RG the
following lightening review is provided. Experts can skip to the
next section.

\subsection{What is the RG?}
 Imagine that you have
some problem in the form of a partition function

\begin{eqnarray} Z(a,b) &=& \int dx \int dy e^{-a(x^2 + y^2)} e^{-b
(x+y)^4}\\ &\equiv &\int dx
\int dy e^{-S(x,y;a,b,..)}\label{Z} \end{eqnarray} where $a$ and $b$ (along
with other  such possible terms) are constant parameters and $S$  is called the action. (In classical statistical mechanics it would be called the energy.)

 The
parameters $b$, $c$ etc., are called {\em couplings} and the
    monomials they multiply are called {\em interactions}. The
    $x^2$ term is called the {\em kinetic} or {\em free-field}
    term,  and $a$, the  coefficient of the kinetic term, has no name and is
    usually set equal to ${1\over 2}$ by rescaling $x$.

 The
average of any $f(x,y)$ is given by

\beq \langle f(x,y)\rangle = {\int dx \int dy f(x,y)e^{-a(x^2 +
y^2)} e^{-b (x+y)^4} \over \int dx \int dy e^{-a(x^2 + y^2)} e^{-b
(x+y)^4}}.\label{avg} \eeq

{\em Suppose we are interested only in functions of
just $x$. } In this case

\begin{eqnarray} \langle f(x)\rangle &=& {\int dx f(x) \int dy e^{-a(x^2 +
y^2)-b (x+y)^4} \over \int dx  \int dy e^{-a(x^2 + y^2)-b
(x+y)^4}}\nonumber \\ &\equiv& {\int dx f(x)  e^{-a'x^2-b' x^4+..} \over \int dx
 e^{-a'x^2- b' x^4 +   ..} } \end{eqnarray}
 where
 \beq
 e^{-a'x^2-b' x^4..}=\int dy e^{-a(x^2 + y^2)-b
(x+y)^4}
\eeq
defines the parameters $a',b'...$etc.
 In the general case we would like to define  an {\em effective action} $S'(x;a',b',c'..) $ by
 \beq
 e^{-S'(x;a',b',c'..)}= \int dy  e^{-S(x,y;a,b,c,..)}
 \eeq
 and the corresponding partition function

\begin{eqnarray}
Z(a', b' ...) &= & \int dx e^{-S'(x;a'b'c'..)}
\end{eqnarray}
  which define the  {\em effective  theory}  for $x$.
   These parameters  $a'$, $b'$ etc., will reproduce exactly the same averages for $x$ as the original  ones $a,b,c..$ did in the presence of $y$.
    {\em This evolution of parameters with the elimination of uninteresting degrees of freedom,
    is  called renormalization. It  has nothing to do with infinities;
    you just saw it happen in a problem with just two variables.}

    Note that even though  $y$ does not appear in the effective theory, its effect has been fully incorporated in the process of integrating it out to generate the renormalized parameters. We  do not say " We are  not interested in $y$, so we will set it equal to zero everywhere it appears", instead we  said, "What theory, involving just $x$, will give the same answers as the original theory that involved $x$ and $y$?"

    The term "Group" arises as follows. Let us say  $y$ is a short
    hand for many variables as is always the case in real life. If
    we eliminate one of them, say $y_1$ which leads to the
    renormalization
     $(a,b,c,..)\to (a',b',c'..)$ and then we eliminate $y_2$  so that now
     $(a',b',c',..)\to (a",b",c"..)$ the net result is
    equivalent to a single  renormalization process in which $(a,b,c,..)\to
    (a",b",c"..)$ under the elimination of $y_1$ and $y_2$. (The
    RG is not a real group since we cannot define a unique inverse.)

\subsection{How is mode elimination actually done?}

Notice that to get the effective theory we need to do a
non-gaussian integral. This can only be done perturbatively. At
the simplest {\em Tree Level}, we simply drop $y$ and find $b'=b$.

In other words, at tree level, the effective action is found by
simply setting the unwanted variables to zero in the original
action.

 At higher orders, we bring down the non-quadratic
exponential and integrate in $y$ term by term and generate
effective interactions for $x$.

 Here is how it is done in our
 illustrative example.
 \begin{eqnarray}
 e^{-S'}&=& e^{-ax^2-bx^4}\int dy
 e^{-ay^2}e^{-b(4xy^3+4x^3y+6x^2y^2+y^4)}\nonumber \\
 \!\! &=& \!\! e^{-ax^2-bx^4}Z_0(a) \!{\int \!\! dy
 e^{-ay^2} \!e^{-b(4xy^3+4x^3y+6x^2y^2+y^4)} \over  Z_0(a)}\nonumber \\
 \!\!\!\!&=& \!\!e^{-ax^2-bx^4} \! Z_0(a) \langle e^{-b(4xy^3+4x^3y+6x^2y^2+y^4)}
 \rangle_{Z_0}
 \end{eqnarray}
 where we have multiplied and divided by $Z_0(a)$,
 \beq
 Z_0(a) = \int dy e^{-ay^2}
 \eeq
 the partition function for a Gaussian action
 $e^{-ay^2}$ and where
 \beq
 \langle e^{-b(4xy^3+4x^3y+6x^2y^2+y^4)} \rangle_{Z_0}\label{eeff}
 \eeq
  stands for the average of the exponential with respect to the partition function $Z_0(a)$.
  Since $Z_0(a)$ is  independent of $x$, we
 will simply ignore it in the effective action  $S'(x;a',b'..)$, without altering  any absolute probability.
 As for Eq. (\ref{eeff}) we invoke the result valid for averages
 over Gaussian actions:
 \beq
\langle e^{-V} \rangle_{Z_0}=e^{-\langle V\rangle - {1 \over 2}
(\langle V^2\rangle  -(\langle V\rangle )^2\rangle \ldots}. \eeq
This is called the cumulant expansion and we see that in  our
example where $V=b(4xy^3+4x^3y+6x^2y^2+y^4)$, the exponent  is  a
power series in $b$ for contributions to $S'$.

To the leading order in $b$ we have  \begin{eqnarray} S'(x;a',b'..)&=& ax^2
+bx^4 +\langle b(4xy^3+4x^3y+6x^2y^2+y^4) \rangle \nonumber \\
&=&ax^2 +bx^4 + 6bx^2 \langle y^2\rangle +\langle y^4\rangle \nonumber \\
&=& ax^2 +bx^4 + 6b{x^2\over 2a}+{3 \over 4a^2}
\end{eqnarray}
because $\langle y \rangle =\langle y^3 \rangle=0$, $\langle
y^2 \rangle={1\over 2a}$ and $ \langle
y^4 \rangle={3\over 4a^2} $.  Thus we have our first RG result for
renormalization: \beq a' =a + {3b\over a}\eeq to leading order in
$b$. Note that we do not care about $x$-independent constants like $  {3\over 4a^2}$.
 For reader who know
 Feynman diagrams, the power series in $b$ can be
 identified with the Feynman graphical expansion for a $\phi^4$ theory (since
 we kept up to quartic terms in the action.)  In the case of the actual $\phi^4$ field  theory itself, when  the momentum cut-off is reduced from $\L$ to $\L/s$ (where $s>1$), $x$ will denote
 low-momentum  modes  $0<k< \L/s$ and $y$ the high-momentum  modes $\L/s<k<\L$,  and
 the loop integration will be  over the range $\L/s<k<\L$.

\subsection{Why do the RG?}

Why do we do this? Because the ultimate effect  of any coupling on
the fate of  $x$ (the variable we care about) is not so apparent when $y$ (the variable in which we have no direct interest) is around, but
surfaces to the top only as we zero in on $x$. For example, we are
going to consider a problem in which $x$ stands for many
low-energy variables and $y$ for many high energy variables.  As we
integrate  out high energy variables and zoom in on the low energy sector, an initially  tiny
coupling can grow in size, or an initially impressive one diminish
into oblivion.

This notion can be made more precise as follows. Consider the
gaussian model in which we have just $a\ne 0$. We have seen that
this value does not change as $y$ is eliminated since $x$ and $y$
do not talk to each other. This is an example of  a {\em fixed
point of the RG} since the coupling that goes in comes out
unchanged by elimination of unwanted variables. \footnote{There
can be more complicated fixed points where this happens despite
the fact that  the $x$'s and $y$'s talk to each other, and the
integrals do not factorize. While we will not encounter  them in
our problem, the proposed strategy applies there as well.} Now
turn on new couplings or "interactions" (corresponding to higher
powers of $x$, $y$ etc.) with coefficients $b$, $c$ and so on. Let
$a'$, $b'$ etc., be the new couplings after $y$ is eliminated.
Let us in addition rescale $x$ so that $x^2$ has the same coefficient as before i.e., $a'=a={1 \over 2} \mbox{say}$. \footnote{We like to keep $a$ fixed because what really matters is the relative size of $a$ and the interaction terms. For example if it turns out that $b'>b$  but also  that $a'>a$, it is not clear the renormalized theory has stronger interactions. Keeping $a$ fixed allows us to compare apples to apples.}
Any of the renormalized couplings
 (still called $b', c',..$), which are are bigger than the initial ones, $b,c,d..$,
are called   {\em relevant} while those that are smaller are called
{\em irrelevant}. This is because in reality $y$ stands for many
variables, and as they are eliminated one by one, the relevant coefficients
 will keep growing  with each elimination and the irrelevant ones will  keep  shrinking  and ultimately disappear.  If a coupling neither grows
not shrinks it is called {\em marginal}. Thus the RG will tell us which couplings really matter in the low energy limit which controls the  nature of the ground state and its low lying  excitations.

There is another excellent reason for using the RG, and that is to
understand the phenomenon of universality in critical phenomena,
an area I will not enter.

 Later we will apply these methods to quantum dots.
For the reader who may  understandably get restless in the interim, here is a glimpse  how we will use RG in the case of dots. First we will begin with all single-particle states in the Hilbert space of dot. Then we will zero in on states  lying within a narrow band  of states within an energy  $E_L$  the Fermi energy $E_F$. (All one needs is that $E_L<<E_F$.) In other words, at this point the variables we called $x$  ($y$ ) are states lying inside  (outside)  this band. For the clean system in the bulk, the result of such a mode elimination is known  \cite{rmp}  to lead to the result that  the system is described by an infinite number of  Landau Fermi Liquid parameters $u_m$, where $m$ is an integer. This process will be discussed in abridged form.  The Universal hamiltonian contains just the $u_0$ term. \footnote{There are actually two sets of $u's$, one for charge and one for spin. The two $u_0$'s are the charging and exchange interactions $u_0$ and $J_0$.} We would like  RG to tell us if and when  we can ignore all other $u_m$'s. To this end we will begin with a theory where only states within $E_T$ of $E_F$ are kept and the starting hamiltonian has all Landau interactions written in terms of the dot eigenfunctions $\phi_{\alpha}$. Then we will ask what happens to them as we eliminate states within $E_T$, getting even closer to $E_F$. To see what happens when we do this (and how we do this) you need to read on!

\section{The problem of interacting fermions}
We will now apply RG to a
  system of nonrelativistic spinless fermions of mass $m$ and momentum $\bK$ in two
space dimensions. The single-particle Hamiltonian is \beq H = {K^2
\over 2m} - \mu \eeq where the chemical potential $\mu$  is
introduced to make sure we have a finite density of particles in
the ground state: all levels  within  the Fermi surface, a circle
defined by \beq {K^{2}_{F} \over 2m} = \mu \eeq
 are now occupied since occupying these levels lowers the ground-state energy.

 In second-quantization we are thus starting with the Hamiltonian
 \beq H_0=\int d^2K
{\psi^{\dag}}( \bK) \left( {K^2-K_{F}^{2}\over 2m}\right)\psi (
\bK) \eeq
where $ \psi^{\dag}( \bK) $ creates a fermion of momentum $\bK$.
 Notice that this system has gapless excitations above
the ground state. You can take an electron just below the Fermi
surface and move it just above, and this costs as little energy as
you please. Such a system will carry a dc current in response to a
dc voltage. An important  question one  asks is if this will be
true when interactions are turned on. For example the system could
develop a gap and become an insulator. Or it could become a
superconductor. How do we decide what the fermionic ground state and low energy excitations
will be, short of solving the problem?

In the noninteracting limit the ground state is simple: it is a single Slater determinant (antisymmetrized wave function)  with all momentum states  below $K_F$ occupied.
If we turn on say,  quartic interactions,  the new ground state will have  pieces in which two fermions from below $K_F$ have been moved to two fermions above $K_F$ keeping the total momentum at  zero.   Each such piece will come with a numerator proportional to the  interaction strength and an  energy denominator equal  to the cost of  creating this particle-hole pair.  Clearly fermions deep in the sea will not take part in this kind of process with any serious likelihood as long as the interactions to be added are weak. Likewise if states far above $K_F$ were removed no one would be the wiser. Thus the fate of the system is going to be decided by states near the Fermi energy. This is
the great difference between this problem and the usual ones in
relativistic field theory and statistical mechanics. Whereas in
the latter examples low energy means small momentum, here it means
small deviations from the Fermi surface. Whereas in these  older
problems we zero in on the origin in momentum space, here we zero
in on a surface. The low energy region is shown in Figure
\ref{annulus}.

We are now going to add interactions and see what they can do.
This is going to involve further reduction of the cut-off. Here we join the discussion of \cite{rmp}.

 Let us begin then with a momentum band of
width $\Lambda$   on either side of the   Fermi surface. (In terms of energy, the band has a width $E_L$, where $L$ stands for Landau, for reasons that will follow. )
\begin{figure}
\centering
\includegraphics[height=2in]{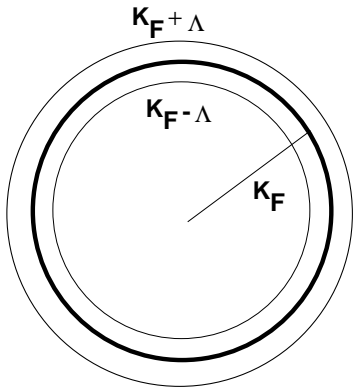}
%
%
\caption{The low energy region for nonrelativistic fermions lies
within the annulus concentric with the Fermi circle. It extends   $\L$ in momentum and $E_L$ in energy from the Fermi circle.  In units where  the Fermi velocity $v_F$ is chosen to be unity, the two are equal. }
\label{annulus}       
\end{figure}

 Let us first learn how to do RG for noninteracting
fermions.
To apply our methods we need to  cast the problem  in the form of
a  path integral. Following any number of sources, say
\cite{rmp} if you want to one-stop shopping,  we obtain the
following expression for the partition function of free fermions:

\beq
Z_0= \int  \left[ d\psi d\overline{\psi}\right]  e^{S_0}\\
\eeq where \beq S_0= \int d^2K \intif d\omega
\overline{\psi}(\omega , \bK) \left(i\omega -
{(K^2-K_{F}^{2})\over 2m}\right)\psi (\omega , \bK)\label{Z0} \eeq
and
\beq
\left[ d\psi d\overline{\psi}\right] =\prod_{\omega, \bK} d\psi (\omega, \bK) d\overline{\psi} (\omega, \bK)
\eeq and $\psi$ and $ \overline{\psi}$ are called Grassmann
variables.
They are really weird objects one gets to love after
some familiarity. There are some rules for doing integrals over
them. For now, I suggest you note only that (i)
$\overline{\psi}(\omega , \bK)$ and $\psi (\omega , \bK) $ are
defined  for each $\omega$ and $\bk$ in the annulus (ii) $Z$ is a
product of Gaussian integrals, with the variables at each $(\omega
, \bK)$ not coupled with those at another. The dedicated reader
can learn more from Ref. \cite{rmp}.

We now adapt this general expression to  a thin  annulus to obtain
\beq
Z_0= \int  \left[ d\psi d\overline{\psi}\right]  e^{S_0}\\
\eeq where
\beq S_0= \int_{0}^{ 2\pi}d\theta \intif d\omega \il dk
\overline{\psi} (i\omega - v_F\ k)\psi \label{Z1}. \eeq We have
approximated as follows: \beq {K^2-K_{F}^{2}\over 2m}\simeq
{K_F\over m}\cdot k=v_F\  k\eeq where $k-K-K_F$ and $v_F$ is the
Fermi velocity, hereafter set equal to unity. Thus $\Lambda$ can
be viewed as a momentum or energy cut-off $E_L$ measured from the Fermi
circle. We have also replaced $KdK$ by $K_F dk$ and absorbed $K_F$
in $\psi$ and $\overline{\psi}$. It will be seen that neglecting
$k$ in relation to $K_F$ is irrelevant in the technical sense.

Let us now perform mode elimination and reduce the cut-off by a
factor $s$. Since this is a gaussian integral, mode elimination
just leads to a multiplicative constant we are not interested in.
So the result is just the same action as above, but with $|k| \le
\Lambda /s$. Consider the following additional transformations:

\begin{eqnarray}
(\omega ', k')                          &=& s(\omega , k)
 \\
(\psi ' (\omega ', k'), \overline{\psi}' (\omega ', k')) &=&\!\!
s^{-3/2} \!\!\left( \! \psi \left({\omega '\over s}, \! {k'\over s}\right)
 , \overline{\psi} \! \left({\omega '\over s}\! , {k'\over s}\!\right)\right).\nonumber \\
 & &
\label{rescale}
\end{eqnarray}

When you  perform the highly recommended exercise of making this
change of variables,  you will find that the action and the phase
space all return to their old values. So what? Recall that our
plan is to evaluate the role of quartic interactions in low energy
physics as we do mode elimination. Now what really matters is not
the absolute size of the quartic term, but its size relative to
the quadratic term. Keeping   the quadratic term identical before
and after the RG action   makes the comparison easy: if the
quartic coupling grows, it is relevant; if it decreases, it is
irrelevant, and if it stays the same it is marginal.

Let us now turn on a generic four-Fermi interaction in
path-integral form: \beq S_4 = \int \overline{\psi}(4)
\overline{\psi}(3) \psi (2)\psi(1) u(4,3,2,1)\label{s4} \eeq where
$\int $ is a shorthand: \beq \int \equiv \prod_{i=1}^{3} \int{d
\theta_{i}}\int_{- \Lambda}^{\Lambda} dk_{i} \intif d\omega_{i}
\eeq

At the tree level, we simply keep the modes within the new
cut-off, rescale fields, frequencies  and momenta , and read off
the new coupling, a highly recommended exercise. We find

\beq u'(k',\omega' , \theta ) = u\left({k'\over s}, {\omega' \over
s}, \theta \right). \label{tree} \eeq

This is the evolution of the {\em coupling function}. To deal with
{\em coupling  constants} with which we are more familiar, we
expand the functions  in a Taylor series (schematic)

\beq u = u_0 + k  u_1  + k^2  u_2 ... \label{series}
\end{equation}
where $k$ stands for all the $k$'s  and $\omega$'s and of course
the $\theta$-s are fixed. An expansion of this kind is possible
since
 couplings in the action are nonsingular in a problem with short range interactions.
If we now make such an expansion and compare coefficients in Eqn.
(\ref{tree}),  we find that  $u_0$ is marginal and the rest are
irrelevant, as is any coupling of more than four fields. For
example

\beq u_{0}^{'}=u_0\ \ \  u_{1}^{'}={u_1 \over s}\ \ \
u_{n}^{'}={u_n \over s^n}\label{rgflow}\eeq

Suppose we start with some cut-off $\Lambda_0$ and an initial
coupling, we shall call $u_n(0)$. Let us now reduce $\Lambda$
continuously as per $\Lambda = \Lambda_0 e^{-t}\equiv \Lambda /s$
so that the change in $u_n$ is continuous in $t$.  The relation
$u_{n}^{'}=u_n/s^n$, Eq. (\ref{rgflow}), can be written as
$u_n(t)=u_n(0)e^{-nt}$,

Sometimes this result is rewritten in terms of  the
$\beta$-function; \beq \beta (t) = {du_n(t)\over dt}= -nu_n(t)\ \
\ \ \mbox{ where  $t=\ln s$.}\label{floweqn}\eeq

If you consider the   $\phi^{4}_{4}$  scalar field theory in four
dimensions you find again an equation like Eq.(\ref{series}) with
the difference that $k$ is measured from the origin and not the
Fermi surface. Thus all we have left to worry about is one {\em
number $u_0$}, the first term in the Taylor series about the
origin, to which the low energy region collapses.  Here the low
energy manifold is a 2-sphere no matter how small $\Lambda$ is and
$u_0$ will inevitably have dependence on the angles on the Fermi
surface:
$$u_0 = u(\theta_1 , \theta_2 , \theta _3 , \theta_4 ) $$

Therefore in this theory we are going to get coupling functions
and not a few coupling constants.

Let us analyze this function. Momentum conservation should allow
us to eliminate one angle. Actually it allows us more because of
the fact that these momenta do not come form the entire plane, but
a  very thin annulus near $K_F$. Look at  Figure \ref{kinematics}.
Assuming that the cutoff has been reduced to the thickness of the
circle in the figure, it is clear that if two points  $\bK_1$ and
$\bK_2$ are chosen from it to represent the incoming lines in a
quartic  coupling, the outgoing ones are forced to be equal to
them (not in their sum, but individually) up to a permutation,
which is irrelevant for spinless fermions. Thus we have in the end
just one function of two angles, and by rotational invariance,
their difference:

\beq u(\theta_1 , \theta_2 , \theta _1 , \theta_2 ) = F(\theta_1 -
\theta_2 )  \equiv  u(\theta ). \eeq About forty years ago Landau
came to the very same conclusion\cite{landau},  that a Fermi system
at low energies would be described by one function defined on the
Fermi surface. He did this without the benefit of the RG and for
that reason, some of the leaps were hard to understand.

\begin{figure} \epsfxsize=3in\epsfysize=2in \hskip
0.0in\epsfbox{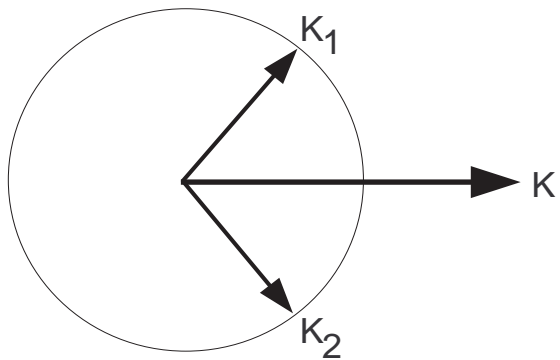}
%
%
\caption{Kinematical reason why momenta are individually conserved
up to a permutation. }
\label{kinematics}       
\end{figure}

Since  $u$ is marginal at tree level, (gets neither big nor small
as the RG process is iterated and modes are eliminated) we have to
go to higher orders in perturbation theory to break the tie, to see
if it ends up being relevant or irrelevant. The answer, given
without proof,  is that it remains marginal to all
orders \cite{rmp}. This is for strange kinematical reasons. A
brief review of how the higher order calculation will be done will
be given when we come to dots.

 Often one writes
\beq u(\theta )=\sum_m u_m \cos (m\theta )\eeq where $u_m$ are the
Landau parameters. The corresponding interaction\footnote{While
this form of $H_L$ will suffice for the dot, in the clean bulk
 system one must allow for small non-forward
scattering since $\Lambda$ is finite. Despite their small measure
these terms are crucial because the Fermi liquid has a  very
singular response at the Fermi surface.} is, in schematic form
with radial integrals over $k$ suppressed,

\beq H_{L}=\sum_{\theta,\theta'} n(\theta ) n(\theta') u_m \cos
\left( m(\theta -\theta')\right)\eeq where $n(\theta )$ is the
number density at angle $\theta$ on the Fermi circle.

Note that if  $u_0 $ is the only non-vanishing Landau coupling, we
get the $u_0N^2$ interaction of $H_U$. If we include spin, there
are spin density-spin density interactions and $J_0$ corresponds
to keeping just the zeroth harmonic. So $H_U$ amounts to keeping
just  the lowest harmonic in the Landau expansion.

We need to ask when and why  $m>0$ terms can be ignored.

\section{RG meets dots}

The first crucial   step towards this goal was taken by  Murthy
and Mathur \cite{qd-us1}. Their strategy  was as follows.
\begin{itemize}
\item  {\bf Step 1:}  Use the clean system RG  described above
   to learn that at low energies the important interactions are the Landau interactions (of which $u_0$ and $J_0$ are a subset).

\item  {\bf Step 2:}   Start at the energy scale $E_T$ and switch to the exact basis states of the
chaotic dot, writing the kinetic term and all the Landau terms  in this
basis. Run the RG by eliminating exact energy eigenstates within
$E_T$.  See what happens to all Landau terms, do they  grow, fall or
stay the same?

\end{itemize}

I will discuss some of the subtleties here, referring you to \cite{gang4} for more details.

A common concern is to  ask if we can  ignore the walls of the dot and use momentum states at high energies. After all,  momentum is not conserved in the dot and what meaning is there to using the Landau interaction written in the momentum basis? The point is this. Consider the collision of two particles in a box. As long as the collision takes place quickly (in terms of the time to cross the box) the incoming and outgoing particles can be labeled by momenta and this label will be useful (conserved) during the collision. \footnote{There will be a  parametrically small   number of cases ( ratio of interaction range to dot size) where collisions take place near  the walls, when this will be wrong.}

However this only brings us down to $E_L$, the energy below which  Landau theory works. Although  $E_L<<E_F$, it is still finite for an infinite system while $E_T\simeq 1/L$ is even smaller for large $L$. In this no man's land between $E_L$ and $E_T$, the flow of couplings is intractable and landau parameters could have evolved from their bulk values.  So what we are saying is this: the universal Hamiltonian has just two of the infinite Landau parameters ($u_0, J_0$) in it.  Let us put in the rest of them with any size, and see what their fate is under further mode elimination.

While this looks like a reasonable plan, it is not clear how it is
going to be executed. There are at least two obvious problems. To
understand them you must understand in more detail how the RG is
done at the one loop level. To illustrate this we use the more
familiar  scalar theory in $d=4$ with the $u_0 \phi^4$ interaction,
where $u_0$, hereafter called just $u$, is marginal and momentum-independent. Suppose we
want the scattering amplitude of four scalar particles. The answer
will be given by a perturbation series, depicted in Fig.
(\ref{fi4}). The second term is an integral over the loop momentum
$K$ that goes from $0\le K\le \Lambda$. The integrand is the
product of the two propagators, which go as $1/K^2$ each at large $K$. (We have
assumed the external momenta are all zero and neglected two other loop diagrams that behave the same way.) The left hand side
corresponds to a physical process and cannot depend on the cut-off
$\Lambda$. It follows $u$ must acquire $\Lambda$-dependence so as
to make the right hand side $\Lambda$-independent. We can find $u
(\L )$ as follows. Suppose $\L$ is reduced by $|d\L |$. The loop
will change by an amount equal to the integral in the region $\L-
|d\L  |<K<\L$,  which is now missing. This missing part must then
come from the first term to keep the physical scattering amplitude
fixed. A simple calculation shows that up to numerical factors
\beq
 du = -u^2 {|d\L |\over \L}\eeq
  so that \beq
 \beta (t) = {du \over dt}=-u^2.\eeq
{\em  In summary,  the change in $u(\L )$ is  given by the loop
terms
 with internal lines  restricted to the modes being eliminated.}

\begin{figure} \centering
\includegraphics[height=1in]{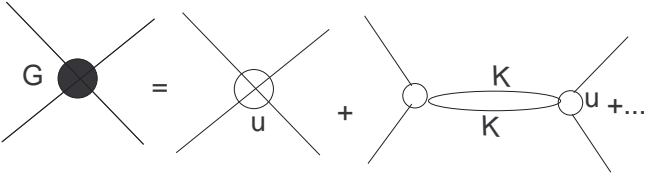}
%
%
\caption{Schematic of scattering amplitude in $\phi^4$ theory. The
solid dot is the full answer and the empty one is the coupling
constant that goes in. The loop is the first  of many that  enter.
}
\label{fi4}       
\end{figure}

We run into the following problems if we try to do the mode
elimination for the dot.

 First, knowledge of the exact
eigenfunctions is needed to even write down the Landau interaction
(the analog of $u$ in the $\phi^4$ example)  in the disordered
basis:
\begin{eqnarray}
V_{\alpha \beta \gamma \delta}&=& \sum\limits_{\bk
\bk'}u (\theta -\theta' ) \left( \phi^{*}_{\alpha}(\bk )
\phi^{*}_{\beta}(\bk' )
-\phi^{*}_{\alpha}(\bk' ) \phi^{*}_{\beta}(\bk )\right) \nonumber\\
& &\times \left( \phi_{\gamma}(\bk' ) \phi_{\delta}(\bk )
-\phi_{\gamma}(\bk ) \phi_{\delta}(\bk' )\right)
\label{V-disorder-basis}
 \end{eqnarray}
where $\bk$ and $\bk'$ take $g$ possible values, $\t$ and $\t'$ are the angles associated with $\bk$ and $\bk'$ and the interaction has been antisymmetrized to mate with the four-Fermi operators $\psi^{\dag}_{\a} \psi^{\dag}_{\b} \psi_{\g}\psi_{\d} $.

Next, additional information is needed on energy levels to do the
higher order (loop) calculation since the propagator for state
$\a$ is $(i\omega -\varepsilon_{\a})^{-1}$. Remarkably it is
possible to overcome ignorance of specific energy levels and
wavefunctions. Here are some, but not all details of how this
comes about.

First consider a {\em specific realization}. We want to reduce the
cut-off by summing over some high energy-states ($y$'s).   If we
write down expression for the one loop flow, four-fold products of
the unknown  wave functions appear at each vertex. At the left
vertex, two of these wave functions correspond to external lines,
which are fixed (and below the new cut-off, i.e., these are the
$x$ variables) while the other two correspond states to be
eliminated and thus summed over. A similar thing happens at the
right vertex. In addition there are the propagators for the two
lines dependent on the single-particle  energies.

Thus what we have here is a product of four  wavefunctions and an
energy denominator summed over   states being eliminated. This
sum will of course vary from dot to dot. Suppose we ignore this
variation and replace the sum by its ensemble average. (The
product of wavefunctions and the energy denominators can be
averaged independently since to leading order in $1/g$, RMT does
not couple them.) The entire average will then be just a function
of just $g$ as in  Eq. (\ref{ficor}).  But what about
sample-to-sample variations?  Here  we invoke  the nice result \cite{qd-us1} that that deviations from the average are down by an
extra power of $g$. {\em Thus the flow is self-averaging, and
every dot will have the same flow as $g\to \infty $. }
\footnote{Experts should note that we are not averaging the
$\beta$-function, it is self-averaging. There may also be concern
that there will not be enough states within $d\L$ to allow the
averaging to work. This too can be handled: there is a  way to
define the $\beta$-function \cite{rmp} in which the loop is first summed over
{\em all} states inside $\L$ and then the $\L$ derivative is
taken.  A simpler
illustration of self-averaging  will be given later when we extend
this result to all orders.}

 Using such a device, these authors found
the remarkable result that the renormalized $V_{\alpha \beta
\gamma \delta}$ is itself equivalent to a Landau interaction but
with  renormalized values of $u_m$ flowing as per  \beq {du_m\over
dt} = -u_m -c u_{m}^{2} \ \ \ \ m\ne 0\eeq where $c$ is
independent of $m$ and of order unity.

Note that $u_0$ does not flow and that  just as in the BCS flow of
the clean system \cite{rmp}, different $m$'s do not mix to this
order. If spin were included $J_0$ wouldn't flow either. This lack
of flow is due to the fact these coefficients multiply operators
that commute with $H_0$, the non-interacting part of the
Hamiltonian, and therefore have no quantum fluctuations.

The flow (for $m \ne 0$) implies that all positive $u_m$'s flow to
zero, as do negative ones with  $u_m>u^*$, the fixed point of the
flow. Thus all points to the right of $u^*$ flow to $H_U$, as
shown in Fig. \ref{bet}. If a modest amount of $u_m, m>0$ of either sign is
turned on at the beginning (when $\L=E_T$), it will renormalize to
zero as we go down in energy. On the other hand if we begin to the
left of $u^*$, we run off to large negative values.

{\em The universal Hamiltonian is thus an RG fixed point with a domain
of attraction of order unity. } To me this is the most satisfactory
explanation of  its success.

\begin{figure} \epsfxsize=2in\epsfysize=.5in \hskip
0.0in\epsfbox{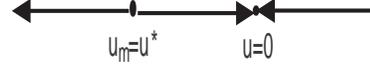}
%
%
\vspace*{1in}
\caption{The flow of the coupling $u_m$ for any $m\ne 0$. The
origin of this flow is the universal Hamiltonian where every
$u_m=0$ except $u_0$ which can have any value. All points to the
right of $u^*$ flow to this. We ask what happens if we begin to
the left. }
\label{bet}       
\end{figure}

\section{The $1/N$ approximation}
So far we have used the RG to understand the low energy behavior of the quantum dot.
 We found that if we begin our analysis with states within $E_T$ of the Fermi energy $E_F$ and all possible Landau interactions $u_m$, and slowly reduce the energy cut-off, we end up with the universal Hamiltonian as the low energy fixed point for all positive $u_m$'s and negative $u_m$'s that are are not more negative than some critical value $u_{m}^{*}$. From this knowledge at very low energies we can conclude in particular that the ground state (lowest of all energies!) is determined by $H_U$  for  $u_m >u_{m}^{*}$ and that at   $u_m =u_{m}^{*}$ the system undergoes a second order phase transition, for that is what a zero of the beta function or fixed point signifies. The RG does not tell us much  about the nature of the new phase, other than that it will be dominated by the $u_m$ that runs off to strong coupling.

 While this is nice, there is clearly room for improvement. In particular,  if the theory could be solved exactly, we would know even more. We would know if the phase transition at $u_m$ of order unity, found in the perturbative, one-loop calculation, really does occur.  We would also   know to what state the system is  driven  after the transition. In other words, if we  can solve a problem exactly, we  do not need the RG. We could,   after obtaining the solution with  some cut-off, easily ask how the coupling is to be modified  if the cut-off is then reduced. We would do this by computing a physical quantity $P(u(\Lambda ), \Lambda ) $ as a function of the input cut-off $\Lambda$ and input coupling $u(\Lambda )$ and then set $dP/d\Lambda =0$ to find $du/d\Lambda $. This would be the exact beta function that one could compare to any perturbative result.

It turns out we can do all  this in our  problem of dots. Now,  any  controlled calculation is   based on some small quantity. If the small quantity vanishes the controlled calculation gives exact results. The most common example is the coupling constant itself, and the exact solution that emerges in the limit of zero coupling is a free theory. However, there are other cases exhibiting  nontrivial behavior, where the coupling is not small but something else is.
For our dots  the small parameter  turns to be   $1/g$  and the exact solution emerges in the limit $g\to \infty$ as
Murthy and
I \cite{qd-us2} found.  Note that  the limit  $g \to \infty $  corresponds to the limit of infinite  dot size. (Unfortunately this does not mean we understand bulk physics, since in this large dot the Thouless band,  over which we have control using RMT,  shrinks to zero as $1/L$.)  What we assume is that what happens in this infinite dot will seen also in large but finite dots, just as one assumes that genuine phase transitions, which are allowed only in infinite systems will be well mimicked in large but finite systems.   We could show that in this limit there is indeed a  transition at a negative coupling of order unity and that to the left of it is a symmetry-broken phase which can be analyzed in some detail.

We showed that one did not have
to rely on RG or perturbation theory in powers of $u$.  Instead the
theory could be solved by saddle point methods for any $u$ thanks to the smallness of $1/g$. The trick was to use a variant of the so called $1/N$ expansion. So let us first get acquainted with the $1/N$ expansion.  I give only a few details here, referring
you to  \cite{gang4}.

Let us begin with the  common situation wherein we expand the answer in  a power series in the coupling $u$, via Feynman diagrams of increasing complexity. This method works provided  the answer can be expanded in a series at $u=0$ and we limit ourselves to weak coupling (since we can typically compute only to some small order in $u$).  There are however cases where we need to  go to all orders in $u$ or  pick up essential singularities before the right physics can be found. Amazingly some of these problems are  tractable thanks to  the  $1/N$ expansion. Here is a brief survey of that trick.

Suppose we have a theory of fermions with an internal isospin or flavor label that runs over $N$ values.   (The method works for bosons as well.)  Let the theory be defined by
the following schematic  path integral: \beq Z=\int \left[ d\psi d
\bar{\psi}\right] e^{\bar{\psi}D\psi+{u\over 2N}(\bar{\psi}\psi)^2} \eeq where $D$
stands for the quadratic kinetic energy term, and the sum over
flavors or integral over space-time is suppressed, so that for
example $$\bar{\psi}\psi= \int d\br d\t \sum_{i}^{N}
\bar{\psi}_{i}\psi_i.$$ The factor of $1/N$ in the quartic term is to ensure that the single sum over the flavor index in the kinetic term has a chance against the double sum in the quartic term. In a theory with Dirac fermions $D= \partial \!\!\!/$ while in a nonrelativistic problem it could be $ \partial /\partial \tau   - \e_{\a}$.

If we now introduce a Hubbard-Stratonovic
field $\sigma$ we can rewrite $Z$ as
\beq
Z=\int \left[ d\psi d
\bar{\psi}\right] d\sigma
e^{\bar{\psi}(D+\sigma
)\psi-N\sigma^2/2u}
\eeq
the correctness of which is readily verified by doing the Gaussian integral over $\sigma$.
If we now use the fact that for {\em each} flavor
\beq
\int \left[ d\psi d
\bar{\psi}\right] e^{\bar{\psi}(D+\sigma
)\psi}=det (D+\sigma
)=e^{Tr \ln (D+\sigma
)}\eeq
and that each flavor gives the same determinant, we obtain
\begin{eqnarray} Z&=&\int \left[ d\psi d
\bar{\psi}\right]  d\sigma
e^{\bar{\psi}(D+\sigma
)\psi-N\sigma^2/2u}\\
&=&\int d\sigma e^{N Tr ln (D+\sigma ) -N\sigma^2/2u}.\label{HS}
\end{eqnarray}

It is now clear that in the limit   $N \to \infty$, we can do the $\sigma $
integral by saddle point. At large and finite $N$, we can compute corrections in a series in $1/N$. However, the complete and exact dependence on $u$ is  obtained to each order in $1/N$. Often just the leading term  at $N=\infty$  captures all the novel physics. A celebrated example is the $1+1$-dimensional Gross-Neveu model \cite{gn},  where there is a coupling constant $u$ and an internal  $O(N)$ isovector index that  runs from $1$ to $N$.   In the limit $N\to \infty$ the saddle point method allows one to show that the $\sigma$ field spontaneously develops a non-zero average (which translates into a mass for the fermion) that goes as  $e^{-1/u}$, a result clearly outside the reach of traditional perturbation theory. Corrections around the saddle point in powers of $1/N$ do not change the main features quoted above. In summary, the small parameter that makes a calculation possible is not $u$ but $1/N$.

We are going to  do the same thing here, with $1/g$ being the small parameter.  You may object  that since
in our problem $D=  \partial /\partial \tau  - \e_{\a}$,  the $g$ different  fermions
are not related by symmetry, and  the appearance of a large number (the analog of $N$)  in front of the Tr ln
is by no means  assured. However, we found that if we went ahead and evaluated   the Tr ln order-by-order in
$\sigma$ and exploited  self-averaging as in the one-loop flow, a large number ($g^2$)  does indeed appear in front of the  action, playing the role of $N$. Here is a glimpse  of how this happens.

Let us first consider just one Landau term with coupling $u_m$, which will simply be called $u$:

\begin{eqnarray} Z&=&\int \left[ d\psi d
\bar{\psi}\right]  d\sigma
e^{\bar{\psi}(D+\sigma
)\psi-\sigma^2/2u}.\label{HS2}
\end{eqnarray}

Next let us make the change $\sigma \to g \sigma$ in Eqn. (\ref{HS2}). While this trivially brings a $g^2$ in front of the $\sigma^2/2u$ term, what is nontrivial is that {\em every term in the Tr ln  (developed in powers of $\sigma$)  will also go as $g^2$ to leading order}. Thus we can write the entire action as $g^2 f(\sigma )$ where $f$ has no $g$ dependence to leading order and the saddle point of $f$ indeed gives exact answers as $g \to \infty$.  I will now show this just for the quadratic term in the Tr ln.
It will then be clear how it works for higher terms.

At a schematic level we can write
\begin{eqnarray}
Tr \ln (D+ g\s)&=& Tr (\ln D + \ln (1 +g D^{-1}\s)\nonumber \\
&=&\!\! C\!\! + \!\! g Tr  D^{-1}\s \!\! -\!\! {g^2\over 2} Tr   D^{-1}\s\ \!\! D^{-1}\s \ldots
\end{eqnarray}
where $C$ is some constant independent of $\s$. When all the indices and details of the Landau interaction are filled for this problem and an $\omega$ integral is done, we find (dropping constants) the following quadratic term
\begin{eqnarray}
 QT &=& \!\! \!\!g^2\!\!\!\!\sum_{\a \b \bk \bk'}\!\! {n_\a -n_\b \over \e_\a-\e_\b}\!\!\left( \s_{1}^{2}\cos m \theta  \cos m\theta' \!\!+ \!\! \s_{2}^{2}\sin m\theta \sin m\theta'  \right) \nonumber \\
 & &\cdot \left(\phi^{*}_{\alpha}(\bk ) \phi^{*}_{\beta}(\bk )\phi^{*}_{\b}(\bk' ) \phi^{}_{\a}(\bk ')\right)
 \end{eqnarray}
 where $\t$ and $\t'$ refer to the angles of $\bk$ and $\bk'$ and $n_{\a}$ and $n_{\b}$ are the Fermi factors,  and $\s_{1}$ and $\s_2$ refer to two components of the $\s$ field. \footnote{ We need two components  because $n(\bk ) n(\bk') \cos m(\t -\t' ) =n(\bk ) n(\bk')( \cos m\t \cos m\t' + \sin m\t \sin m\t' )$ is a sum of two terms and each needs to be factorized with its own Hubbard-Stratonovic field. We will refer to the two-component vector as $\sigmab$.}

 We will replace this term by its ensemble average, since   deviations from the average are down by an extra power of $1/g$ and thus  ignorable as $g \to \infty$  \cite{qd-us1}. As for the average, recall that  to leading order in $1/g$,  averages of energy levels and wavefunctions factorize. Using
 \beq
 \langle \phi^{*}_{\alpha}(\bk_1 ) \phi^{}_{\beta}(\bk_2 )\phi^{*}_{\b}(\bk_3 ) \phi^{}_{\a}(\bk_4)\rangle ={\delta_{\bk_1 \bk_4}\over g}{\delta_{\bk_2 \bk_3}\over g}+ {\cal O}(1/g^3)
 \eeq
 for our case where $\bk_1 = \bk_2= \bk$  and $\bk_3 = \bk_4= \bk'$ and summing over the $g$ values of $\bk$ and $\bk'$ we find this average over wavefunctions goes as $1/g$. (We are not getting into details like a factor of $1/2$ coming from the $\cos^2 m\t$ and $\sin^2 m \t$ in the momentum sums.)

 As for the sum over energy denominators, consider one of the two similar possibilities, where $n_{\a}=1$ and $n_{\b}=0$ so that $\e_{\a}$ ranges from $-g\delta /2$ and $0$ while $\e_{\b}$ ranges from $0$ to $g\delta /2$.  Replacing  the sum by an integral over density of states $1/\delta$ (valid for the ensemble average) we find an integral of the form
 \beq
  {1 \over \d^2} \int_{0}^{g\d /2}  \int_{0}^{g\d /2}{dE dE' \over E+E'}\simeq {g\over \d }\eeq
  up to constants, so that that
  \beq
  \langle QT \rangle \simeq {g^2 \over \d}\s^2
  \eeq
  where $\s^2= \s_{1}^{2}+\s_{2}^{2}$. If we consider higher powers of $\s$ we again find that they go as ${g^2 \over \d}\s^{2n}$.  Had we not scaled $\s$ initially, we would have found  that each power of the unscaled $\s$ was accompanied by a  different power of $g$ and that no  large $g$ limit  emerged.

  It can be shown that had we  included all $u_m$'s simultaneously,  they would  not have interfered at the quadratic level. Since this is the term that controls  symmetry breaking, we get the right picture taking  just one $u_m$ at a time.

For the experts I mention that since we have a large $N$ theory
here, it follows  as in all large $N$ theories, that  the one-loop
flow and the new fixed point at strong coupling are parts of the
final theory. However the exact location of the critical point
cannot be predicted, as pointed out to us by Professor Piet
Brower. The reason is that the Landau couplings $u_m$ are defined
at a scale $E_L$ much higher than $E_T$ (but much smaller than
$E_F$) and their flow till we come down to $E_T$, where our
analysis begins, is not within the regime we can control. In other
words we can locate $u^*$ in terms of what couplings we begin with
at $E_T$, but these are  the Landau parameters renormalized in a
nonuniversal way as we come down from
 $E_L$ to $E_T$.

 What is the nature of the  state for $u_m\le
u_m^*$?

In the strong coupling region $\sigmab$  acquires an expectation
value in the ground state. The dynamics of the fermions is
affected by this variable in many ways: quasi-particle widths
become  broad very quickly above the Fermi energy, $\Delta$ ( spacings
in $V_g$ between successive peaks)
has occasionally very large values and can even be negative,
\footnote{How can the cost of adding one particle be negative
(after removing the charging energy)? The answer is that adding a
new particle sometimes lowers the energy of the collective
variable which has a life of its own. However, if we turn a blind
eye to it and attribute all the energy to the single particle
excitations, $\Delta$ can be negative.} and the system behaves
like one with broken time-reversal symmetry if $m$ is
odd\cite{gang4}. One example of the latter is as follows. Suppose we turn on a weak magnetic field $B$. All quantities- peak positions,  spacing -  will vary {\em linearly} in $B$ and not quadratically as in a time-reversal invariant system. \footnote{ This result breaks down at exponentially small $B$, a region we can safely ignore since the temperature in any realistic experiment will be much higer.}

Long ago Pomeranchuk \cite{pomeranchuk} found that if a Landau
parameter $u_m$   of a pure system exceeded a certain value, the Fermi
surface underwent a shape transformation from a circle  to a
non-rotationally invariant form. Recently this transition has
received a lot of attention\cite{varma,oganesyan} The transition
in question is a disordered version of the same. Details are given
in Refs. \cite{qd-us2}, \cite{gang4}.

Details aside, there is another very interesting point: even if
the coupling does not take us over to the strong-coupling phase,
we can see vestiges of the critical point $u_{m}^{*}$ and
associated critical phenomena. This is a general feature of many
{\em quantum critical points}\cite{critical-fan1,critical-fan2}, i.e., points
like $u_{m}^{*}$,  where as a variable in a Hamiltonian is changed, ground state of
the system undergoes a phase transition  (in contrast to transitions wherein
temperature $T$ is the control parameter).

\begin{figure} \centering
\includegraphics[height=2in]{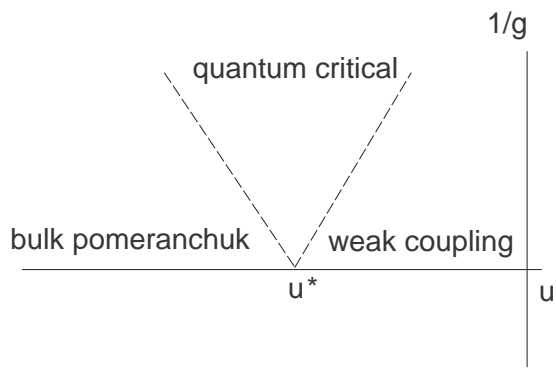}
%
%
\caption{The phase diagram in the $u-{1 \over g}$ plane. The
actual phase transition takes place on the line $1/g=0$. However
it can be perceived at finite $g$ if one is within the $V$-shaped
quantum critical regime. In typical cases, the $1/g$ axis is
replaced by the $T$ axis, where $T$ is the temperature. There a
phase transition of quantum mechanical origin at $T=0$ influences
the quantum critical regime at finite $T$.  }
\label{fan}       
\end{figure}

Figure \ref{fan} shows what  happens in a generic situation. On
the $x$-axis a variable ($u_m$ in our case ) along which the
quantum phase transition occurs. Along $y$ is measured a new
variable, usually temperature $T$. Let us consider that case
first. If we move from right to left at some value of $T$, we will
first encounter physics of the weak-coupling phase determined by
the weak-coupling fixed point at the origin. Then we cross into
the {\em critical fan} (delineated by the $V$-shaped dotted
lines), where the physics is controlled by the quantum critical
point. In other words we can tell there is a critical point on the
$x$ -axis without actually traversing it. As we move further to
the left, we reach the strongly-coupled symmetry-broken phase,
with a non-zero order parameter.

In our problem, $1/g^2$ plays the role of
$T$ since  $g^2$ stands in front of the effective action for
$\sigmab$. (Here $g$ also enters at a subdominant level inside the
action, which makes it hard to predict the exact shape of the
critical fan.) The bottom line is that we can see the critical
point at finite $1/g$. In addition one can also raise the actual temperature $T$
or bias voltage to see the critical fan.

Subsequent work has shown, in more familiar examples than Landau
interactions,  that the general picture depicted here is true in
the large $g$ limit: upon adding sufficiently strong interactions
 the Universal
Hamiltonian gives way to other descriptions with broken
symmetry\cite{browergm}.

It was mentioned earlier    that the critical coupling  $u^*$ (a
nonuniversal quantity)  cannot be reliably predicted in the large
$g$ limit. It has become clear from numerical work  \cite{brouwer} that it coincides
with the bulk coupling for the Pomeranchuk transition. In other
words, when we cross over  to the left $u^*$,  the size of the
order parameter very rapidly grows from the mesoscopic scale of
order $E_T$ to something of order the Fermi energy $E_F$. However the
physics in the critical fan as well as the weak coupling side is
as described by the  RMT+RG analysis. The strong coupling side has
to be reworked from scratch since the Fermi surface assumed in the
RG that came down to $E_T$ is has suffered huge deformations (in
the scale of $E_F$).

\section{Summary and conclusions}

Our goal was to understand the transport properties of a quantum dot- an island that  electrons could tunnel  on to and tunnel  out of. As one varies the gate voltage $V_g$ between the leads and the dot, the conductance $G$ exhibits  isolated peaks of varying location and height. What we wanted was a {\em statistical description} of these features as exhibited by an ensemble of similar dots. The dot was assumed to be so irregular and the classical motion so chaotic, that the only conserved quantity was energy.

To get warmed up, we asked how we would go about addressing the problem if electron-electron interactions could be neglected. It was seen that peak heights  and positions  could be determined from  the wavefunctions and energy levels. These in turn could be determined by RMT given just the mean level spacing. In the non-interacting theory, $P(\Delta )$, the distribution of differences in $V_g$ between successive peaks,  was the same as $P(\d )$, the distribution of spacings between successive levels in the dot.

Actual comparison of these two distributions showed a clear disagreement, which represented a failure, not of RMT, but  of the assumption of non-interacting electrons. So interactions had to be included.

The coulomb  interaction implied that adding an electron to the dot  would cost not just the gap to the next empty level, but an additional energy due to repulsion by the $N$ electrons already in the dot. This charging energy could be accounted for  by adding a term $u_0 N^2$ to the second quantized Hamiltonian. In the presence of spin,  we also needed to add a term $-J_0 S^2$ to represent the exchange interaction. The final result was $H_U$, the universal Hamiltonian. While the success of this model is unquestioned at moderate values of interaction strength, some arguments for why this had to be the right answer, and why other interactions could be neglected because their  ensemble averages vanished were not persuasive. We asked if there was there a better way to understand the success of $H_U$.

It was pointed out that the RG was such a way since it  offered an unbiased procedure for  determining  which interactions were really  important in deciding low energy properties like  the ground state and its low energy excitations. In the  RG approach  one divided the variables into two sets: $x$, which we cared about,  and $y$,  which we we did not care about. In our example $x$ was the low energy region (near the Fermi energy) and $y$  everything else. One then eliminated or integrated out the $y$'s to  obtain an effective theory of just $x$, {\em which gave the same answers in the $x$-domain as the original one.} In this process some initially very impressive couplings could fade into oblivion (irrelevant) while tiny ones could  grow in size (relevant) and some could remain fixed (marginal). In any event we could see which couplings really mattered.
We saw that while the RG concept itself was non-perturbative, mode elimination was typically done perturbatively  in the interaction.

The application of RG to our problem required a two-stage process as developed by Murthy and Mathur. First one ignores disorder and finite size  and  eliminates high energy modes outside the  Landau band, a region of width   $E_L$ measured from the Fermi surface. Here we know from past RG work on clean bulk systems  that we must invariably end up with  the Landau interaction $u(\t )$.  But we are not done yet. Although the Landau scale $E_L$ is  much smaller than $E_F$, the Fermi energy,  it does not vanish for infinite system size. So $E_T$, which vanishes as $1/L$, lies  even closer to $E_F$. So we need to renormalize down from $E_L$ to $E_T$. During this process the Landau parameters could renormalize in a way we cannot determine.  (Though in a clean system $u_m$ are strictly marginal, once we approach $E_T$ and take disorder seriously, a nonzero flow is guaranteed. )  So one begins  at  $E_T$ by   writing  the (renormalized) Landau interaction in the disordered single-particle basis $\a$ and eliminating  the $\a$ states  to determine the fate of the Landau interactions. It was found that only $u_0$ and $J_0$, the zeroth harmonics on the Fermi circle of the Landau interactions (for charge and spin densities), survived at the lowest energies if the starting value of $u_m$ was  either positive  or not below a negative  coupling $u_{m}^{*}$, of order unity. Thus the low energy   fixed point for this range of initial coupling was just $H_U$, the universal Hamiltonian. In addition to providing this justification of the emergence of $H_U$, the calculation also suggested that for $u_m<u_{m}^{*}$  the system underwent a phase transition in the $g \to \infty $ limit. Before discussing the phase transition,  let us recall how the calculation of the $\b$-function was done.

In any interacting field theory  with some coupling $u$, one computes a physical quantity like a scattering amplitude in a power series in $u$, starting with $u$ itself, followed by loop diagrams of increasing complexity. These loops involve momentum or energy sums (or integrals) up to some cut-off $\L$. Clearly if the sum of all these terms has to be independent of $\L$ (as is the physical scattering amplitude) then $u$ itself must become $u(\L )$ and vary with $\L$ in such a way as to keep the series as a whole $\L$-independent.  Conversely it is possible to determine $u(\L )$ by drawing diagrams to some order in $u$ making this demand.
 In our problem, the computation of $\b (u)=-\L du/d\L$ for any one specific dot required knowledge of the wave-functions and energy levels lying within $E_T$.  But  thanks to self-averaging, one could replace the $\b$- function for the given dot by its ensemble average.  The zeroes of   this $\b$-function   are what showed $H_U$ to be a fixed point (at the origin) and  $u_{m}^{*}$ to be the the critical point for the phase transition.

 This clever calculation was nonetheless a weak-coupling analysis  predicting  a phase transition at strong coupling. Was the transition real and if so, what was on the strong coupling  side?  It was here that the $1/N$ technique  came in. In the large $g$ limit one could show that all physics could be extracted in a saddle point calculation of the $1/N$ type for any coupling $u_m$. In contrast to theories where there were $N$ equivalent species, here we had $g$ fermions with different energies and matrix elements. However, thanks to  disorder self-averaging, one could pull out a $g^2$ in front of the action for the Hubbard-Stratonovic field $\sigmab$. The saddle point theory confirmed the fixed point nature of $H_U$ for $u_{m}^{*}<u_m<\infty$ and the transition at $ u_{m}^{*}$. Furthermore we could see beyond the transition to the other side: here
 $\sigmab$ acquired an average (a disordered version of the Pomeranchuk transition in clean systems) and produced many attendant consequences like time-reversal breaking.  However the "exact" critical point $u_m=u_{m}^{*}$ of  this  calculation was not a directly measurable quantity  since the saddle point theory had as its input, not the Landau interaction defined at $E_L$, but what it had evolved into,  between $E_L$ and $E_T$. This was a  no man's land where   disorder was too strong to be ignored, but not strong enough to use RMT since we were not  within $E_T$ making the flow intractable. However clever arguments of Adam {\em et a}  show  that the transition occurs at the bulk critical value of $u_m=1$.

Since the phase transition occurs only at infinite $g$ (a finite system  always has finite $g$ and cannot have a transition), it might seem that our study of it was academic. This is not so, and the reason is the same as in quantum critical phenomena. Recall that there, a quantum phase transition at $T=0$ as a function of some coupling $u=u^*$ can be felt even at $T>0$ inside a $V$-shaped region called the quantum critical region. Here $g^2$ plays the role of $1/T$  as a prefactor in the action. Thus we can see the effects of the critical point even at finite $g$ and over a wide range of coupling.

I began with the ominous  remark that  three obstacles are ganged up here: randomness, strong interactions and finite size. Yet they ended up being benign: randomness and finite size led to a finite Thouless band within which we could use RMT and ply our trade, while strong interactions led to an interesting phase transition.
As for the RG, it had to be invoked in the first stage as  we came down to $E_L$ using the clean system RG to end up with the Landau interactions.  At this point we could either use  perturbative self-averaged RG inside $E_T$ or better still, use the self-averaged $1/N$ method  to solve the model by saddle point.

This colloquium has emphasized what I find most beautiful about this problem: the confluence  of physical complications and the  interplay  of diverse techniques that lead to a solution. Of necessity it has been sparse on phenomenology.  However, armed with the ideas explained here you are ready to remedy this,  following any number of the excellent references mentioned in the text.

  It will be very interesting if experimentalists unearthed the phenomena chronicled here by
 studying dots with strongly interacting electrons, a possibility
more readily realized   than in  the bulk since electron
density in dots can be controlled by gates. Stay tuned for
these  results.

\section*{Acknowledgments}

I am   grateful to the National science Foundation for grant
DMR-0354517 that made this research possible. I thank my constant
collaborator Ganpathy Murthy for so many shared insights including
those pertaining to this article.

\bibliographystyle{apsrmp}


\newpage

\end{document}